\documentclass{article}
\usepackage{spconf,amsmath,graphicx,hyperref}

\usepackage{cite}
\usepackage{amsmath,amssymb,amsfonts}
\usepackage{algorithmic}
\usepackage{textcomp}
\usepackage{xcolor}
\usepackage{booktabs}
\usepackage{subcaption}

\usepackage{circuitikz}
\usepackage[nolist]{acronym}
\usepackage{tikz}
\usepackage{url}

\newcommand{\precursor}{\mathtt{P}}

\newcommand{\cleanSignal}{\mathbf{x}}
\newcommand{\degradedSignal}{\mathbf{y}}

\newcommand{\numSamples}{T}
\newcommand{\numFrames}{F}

\newcommand{\degradedSet}{\mathcal{Y}}

\newcommand{\Figure}[1]{Figure~\ref{#1}}

\title{A Robust, Explainable, and Efficient Deep Metric for Speech and General Audio Quality}
\name{Konstantin Schmidt, Pablo M. Delgado, and Andreas Brendel \sthanks{Andreas Brendel has been supported by the Free State of Bavaria by the DSgenAI project.}}
\address{\textit{Fraunhofer Institute of Integrated Circuits (IIS),}
Erlangen, Germany\\ 
}

\begin{acronym}
    \acro{FM}[FM]{Foundation Model}
	\acro{SAQM}[SAQM]{Speech and Audio Quality Metric}
    \acro{SOTA}[SOTA]{state-of-the-art}
    \acro{MOS}[MOS]{Mean Opinion Scale}
    \acro{PEAQ}[PEAQ]{Perceptual Evaluation of Audio Quality}
    \acrodef{DNN}[DNN]{Deep Neural Network}
\end{acronym}

\begin{document}
\ninept
\maketitle
\begin{abstract}
We present a robust \ac{SAQM} that merges features of traditional quality metrics with a CNN trained on a large amount of data using contrastive learning methods.
To overcome the problem of limited amounts of human-labeled training data, we propose two pretraining strategies relying on pseudo-labels. One uses an established objective metric PEAQ-2f (based on ITU-R BS.1387) as surrogate loss, the other is based on contrasting a wide range of coding modes by a rule-of-thumb ordering based on expected perceived quality.
As a key novelty, our approach lets us generate saliency maps by propagating gradients to identify which feature has the greatest impact on the final metric outcome and enables our system to provide insights regarding the type and extent of distortions present in the tested audio signal.
The resulting \ac{SAQM} operates at low computational complexity, performs comparably to \ac{SOTA} metrics on speech while still performing well on general audio content.
\end{abstract}
\section{Introduction}
In this work, we present the design of a robust \acf{SAQM} for monitoring perceived quality of coded audio content with explainable model predictions. 
State-of-the-art systems can be distinguished into first, established methods that built upon hand-crafted, intricate rule-based features that are perceptually motivated \cite{PESQ, PEAQ} and second, systems that build upon \acp{DNN} like audio foundation models \cite{NOMAD, SCOREQ, DeePAQ}.
Systems falling into the first category exhibit strong correlation to subjective ratings for specific degradations caused by processes such as perceptual coding, speech enhancement or blind-source separation  \cite{PESQ, PEAQ, DelgadoPEAQ}. 
However, these traditional models often show low performance and struggle with reliability when evaluating modern or complex impairments, including additive noise, reverberation, or bandwidth-extended (BWE) speech and music \cite{reddy2022dnsmos,abel2016instrumental,delgado2023improved}.

Systems from the second category are primarily purely data-driven \ac{DNN}-based models, which have outperformed traditional systems in assessing speech affected by convolutive and additive noise \cite{reddy2022dnsmos}, predicting quality in lossy transmission channels \cite{affonso2017speech}, and evaluating BWE-processed signals \cite{abel2016instrumental}.

Interestingly, by just calculating statistical metrics between the embeddings of original and degraded signals in the embedding space of audio foundation models, good correlation with human ratings can be achieved even without further training\cite{kilgour19_interspeech,chung_kad_2025,gui_adapting_2024, LPIPS}. However, these foundation models are mostly trained with speech signals and their generalization capabilities to general audio content is limited. Also most of these models are trained with 16 kHz sampled signals which makes them blind for high frequency degradations. 
Lately, systems based on contrastive learning outperform above mentioned models by a significant margin on certain datasets \cite{NOMAD, SCOREQ, UTMOS, jiang2025deepaqperceptualaudioquality}. 
Noteworthy, the approach presented in \cite{SCOREQ} introduces a triplet loss function adapted for contrastive regression. The primary benefit of this approach is learning a robust quality manifold by contrasting triplets (anchor, positive, negative) based on \ac{MOS} distance, thereby ensuring that the embedding space distance reflects the required quality ordering, unlike traditional L2 loss embeddings which tend to capture degradation type instead of quality. This results in reported significant and consistent performance improvements and superior generalization across out-of-domain and out-of-distribution speech datasets.

A major challenge for these pure learning-based metrics is the scarcity of high-quality human-annotated subjective data required for supervised training, which can limit generalization ability \cite{manocha2021noresqa}. This has recently been investigated for assessing the quality of combined speech and music signals in \cite{wisnu2025haaqi, DeePAQ}. 
We ally this problem by proposing two pretraining strategies that do not rely on human annotated data.

Moreover, \ac{DNN}-based quality metrics are often viewed as an incomprehensible black box, leading to issues with interpretability and tractability \cite{Bommasani2021FoundationModels}. This lack of transparency hinders the ability to precisely diagnose the cause or locality of quality degradation, aspects that are often better facilitated by the inherent interpretability of detailed, rule-based perceptual models.

Established methods are usually intrusive or full reference systems, meaning that they rely on the coded and the original signal for calculating a metric. Modern \ac{DNN}-based systems do not necessarily need a reference. This makes them applicable to estimating the quality of artificially generated (non waveform preserving) content. However we will show that their performance on general audio content drops since these signals and degradations were never seen during training.

In this work, we demonstrate that features from the ITU-R BS.1387 perceptual model \ac{PEAQ} \cite{PEAQ} can significantly reduce the extensive parameter space of general audio signals while still maintaining the ability to detect degradations. Because the \ac{PEAQ} features we use measure certain auditory events related to the perception of specific artifacts, we can compute saliency maps to explain which features contribute most to the final \ac{SAQM} score. This means that our system can provide feedback about the type, position and amount of distortion present in the audio signal.

To summarize our contribution: we propose a \ac{SAQM} with the following key aspects:
\begin{itemize}
    \item Strong and robust performance: high Pearson correlation across diverse datasets, covering both speech and general audio with a wide range of distortions.
    \item Explainability: feedback on the type and locality of distortions.
    \item Low computational complexity compared to systems based on foundation models.
\end{itemize}

\section{System Description}
\begin{figure}[htbp]
    \centering
    \includegraphics[trim={0.8cm 0.3cm 0.8cm 0cm}, clip, width=\columnwidth]{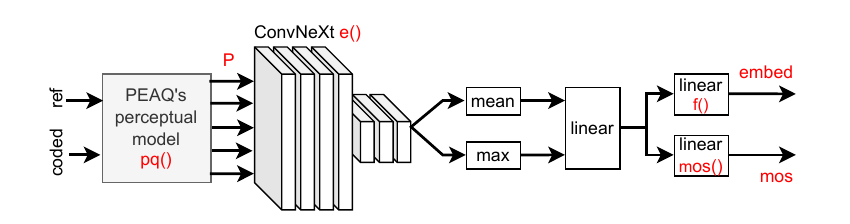}
    \caption{Proposed system consisting of precursor calculation $\mathtt{pq}(\cdot)$, CNN $e(\cdot)$, embedding layer $f(\cdot)$ and prediction head $\mathrm{mos}(\cdot)$.}
    \label{fig:block_diagram}
\end{figure}

Our proposed system, depicted in \Figure{fig:block_diagram}, extends the distortion-to-quality mapping of PEAQ with a contrastive learning approach. 
We assume that a clean speech or audio signal $\cleanSignal\in\mathbb{R}^\numSamples$, where $\numSamples$ denotes the signal length in samples, undergoes a degradation (e.g. coding), yielding a degraded signal $\degradedSignal\in\mathbb{R}^\numSamples$.

The PEAQ Basic model \cite{PEAQ} extracts distortion metrics by comparing perceptual representations of $\cleanSignal$ and $\degradedSignal$ that serve as the instantaneous, frame-by-frame psycho-acoustic measurements that quantify quality degradation in different aspects per perceptual frequency band for all $\numFrames$ signal frames. We name these time-frequency representations \textit{precursors} \footnote{\label{noteKabal} As implemented in \cite{Kabal}, available at: \url{https://www.mmsp.ece.mcgill.ca/Documents/Software/}}

\begin{equation}
    \precursor := \mathtt{pq}(\cleanSignal,\degradedSignal)\in \mathbb{R}^{1199 \times \numFrames},
\end{equation}
where each of the $11$ precursors has a frequency resolution of $109$ which are concatenated in the frequency domain yielding $1199$ features.
These precursors measure signal loudness, noise-loudness and bandwidth of reference and test signal; tonal error in the correlation power spectrum and instantaneous modulation differences. Please see \cite{Kabal} for more details.
For calculating $\precursor$, the input signals are time-aligned, normalized to a playback level of 92 dB SPL, and resampled to 48 kHz to match PEAQ's perceptual model’s calibrated constants.

The precursors $\precursor$ are fed into a DNN $e(\cdot)$ consisting of a stack of convolutional layers based on the ConvNeXt architecture \cite{ConvNeXt}, followed by a parallel max and mean pooling over the time dimension and a linear layer that combines the pooled results. The output $e(\precursor)$ is fed into two projection heads $f(\cdot)$ and $\mathrm{mos}(\cdot)$ which are implemented as linear layers similar to SCOREQ \cite{SCOREQ}. The complete mapping of input to embedding is denoted by $F(\cdot,\cdot) = f(e(\mathtt{pq}(\cdot,\cdot)))$.



%
\subsection{Weakly-supervised Training}\label{weak_train}
To allay the limited amount of annotated training data we propose two weakly-supervised learning strategies that do not need any human annotations. 
Since our main focus is monitoring the quality of speech and audio coding, we use the xHE-AAC codec \cite{USAC} operating at various bitrates, ranging from 32 to 80 kbps (mono), and core mode settings to generate the training data. 
The xHE-AAC codec is configured to operate in the time and in the frequency domain as well as with and without parametric coding tools such as Temporal Noise Shaping (TNS) \cite{herreTNS}, Spectral Band Replication (SBR) \cite{Ekstrand2002BANDWIDTHEO} and Intelligent Gap Filling (IGF) \cite{disch2016intelligent}. The time-domain mode utilizes a speech codec akin to Adaptive Multi-Rate Wideband AMR-WB \cite{AMR_WB}, whereas the frequency-domain mode encodes spectral lines from an Modified Discrete Cosine Transform (MDCT) transform.

\textbf{Pretraining 1:} 
First, we use pseudo labels as quality scores $m$. 
For the general audio part of the train set we use PEAQ-2f \cite{PEAQ} (divided by 20 to match the MOS scale) and for the speech part we use averaged listening test results for each codec configuration.
Here, we assume that although these labels may not perfectly align with human ratings, they are generally capable of ranking degradations based on their intensity. 
We use the batch-all triplet loss \cite{SCOREQ} to generate triplets indexed by $i,j,k$ of all files in the batch
\begin{multline}\label{eq:triplet_loss}
    \mathcal{L}_{\text{triplet}} := \sum_{i,j,k=1}^{N} M(i,j,k)  \max\bigg(0, \|F(\cleanSignal_i,\degradedSignal_i) - F(\cleanSignal_j,\degradedSignal_j)\|^2\\ -\|F(\cleanSignal_i,\degradedSignal_i) - F(\cleanSignal_k,\degradedSignal_k)\|^2 + g\bigg),
\end{multline}
wit the adaptive margin $g:= \frac{| m_i - m_j | - | m_i - m_k |}{4}$ as in \cite{SCOREQ}, $m$ quantifies percetpual quality. 
The binary mask $M(i,j,k)$ selects the valid combinations of items depending on the absolute difference of perceptual quality $m$
\begin{equation}
    M(i,j,k) = 
    \begin{cases}
        1 & \text{if}\ | m_i - m_j | < | m_i - m_k |\ \text{for}\ i \ne j \neq k, \\
        0 & \text{otherwise}.
    \end{cases}
\end{equation}

\textbf{Pretraining 2} The next training iterations are based on the assumption that some configurations of the xHE-AAC codec are on average better than others. E.g. significant higher bitrates are mostly better than lower ones when keeping other codec configurations fixed. 
To increase the number of contrastable configurations, we also let the xHE-AAC codec run in ill-posed configurations. E.g. we assume that the codec operating in short blocks only performs at the same bitrate always worse than when operating with the hand-tuned block switching decision typically used in xHE-AAC.
Here, the loss is evaluated on a train set comprising $I$ tuples $\degradedSet = \{(\degradedSignal^i_1,\degradedSignal^i_2,\degradedSignal^i_3)\}_{i=1}^I$ of contrastable configurations where coding mode of $\degradedSignal^i_1$ has better perceived quality than the mode of $\degradedSignal^i_2$ and mode $\degradedSignal^i_2$ is better than mode $\degradedSignal^i_3$ \cite{FaceNet}
\begin{multline}
    \mathcal{L}_{\text{pretrain}} := \max\bigg(0, \|F(\cleanSignal,\degradedSignal_1) - F(\cleanSignal,\degradedSignal_2 )\|^2\\ -\|F(\cleanSignal,\degradedSignal_1) - F(\cleanSignal,\degradedSignal_3)\|^2 + g\bigg).
\end{multline}

\textbf{Subsequent Training}
For the subsequent training iterations, the system is fine-tuned on human-annotated data. We use the loss in Eq. \ref{eq:triplet_loss}, replacing the ratings $m$ with human ratings.

\section{Evaluation}
\subsection{Experimental Setup}
The CNN we use is based on the ConvNeXt architecture \cite{ConvNeXt}, where we replace the 2-dimensional convolutions by 1-dimensional convolutions and use 4 blocks of dimensions [517, 256, 256, 256]. To save complexity in the first layer,  we split the 1199 feature channels corresponding to MOV dimensions into 11 groups. For this reason the first layer needs 517 channels.
We apply batch normalization before the initial convolutional layer to normalize the PEAQ precursors.
To stabilize training and improve generalization, we maintain an exponential moving average (EMA) of the model parameters \cite{moralesEMA}. 
The complexity and memory requirements of our model are compared to other models are:\\
\begin{table}[t!]
    \begin{tabular}{l|ccc}
 & ConvNeXt & SCOREQ & DeepAQ \\ 
\midrule
multiply-adds [giga]       & 0.039 & 20 & 11 \\ 
DNN parameters [million] & 20    & 95 & 97 
\end{tabular}
\caption{Complexity and parameters of proposed and \ac{SOTA} nodels}
\vspace{-2mm}
\end{table}
For the pretraining, we use 55 different codec configurations. The audio material consists of roughly 50 \% speech content taken from the VCTK dataset and the other half taken from popular music recordings. In total we use about 5000 hours of training data. 
We train the model for 100 epochs by randomly switching between the training strategies described in Sec. \ref{weak_train}.
The training batches consist of randomly cropped 3-second audio snippets with PEAQ-2f values computed on the fly.
For the final training iterations, we use the training set from the NISQA Corpus \cite{NISQA_DS}. This Corpus includes more than 14,000 speech samples with degradations from speech codecs, packet-loss, background noise as well as recordings from mobile devices using Zoom, Skype, WhatsApp etc.
Each file was rated by human listeners in a P.800 test (without reference), yielding more than 97,000 human ratings in total.

For evaluation, we use 12 datasets with human ratings that are publicly available. These datasets include the 4 subsets from the NISQA Corpus: \textit{NISQA-VAL-SIM}, \textit{NISQA-VAL-LIVE}, \textit{NISQA-LIVETALK}, \textit{NISQA-FOR} and \textit{NISQA-P501}, which were given a \ac{MOS} in a P.800 methodology where subjects don't get to listen to the reference.
The \textit{NISQA-FOR} and \textit{NISQA-P501} subsets include packet-loss items in which frames are either dropped or additionally inserted, leading to a segment-wise variable delay. Since the precursors employed in the proposed model cannot cope with such items, we excluded all packet-loss items from these subsets and delay-aligned the remaining files.

The Open Dataset of Audio Quality (\textit{ODAQ}) \cite{ODAQ_dataset} comprises 240 audio samples, each evaluated by 26 listeners and processed to obtain various quality levels across six distortion classes: Low-Pass, Pre-Echoes, Spectral Holes, Tonality Mismatch, Unmasked Noise, and Dialogue Enhancement artifacts.

The \textit{IgorC96} dataset \cite{IgorC96Multiformat} compares Opus, AAC, and Ogg Vorbis at 96 kbps against mp3 at 128 kbps. It includes 40 items, mainly music.

The two MPEG USAC Verification Tests \cite{USAC_DS} contain items coded with the Unified-Speech-and-Audio-Coding codec (USAC), AMR-WB+ and HE-AAC v2 at different bitrates. \textit{USAC-1} contains mono items at low bitrates (8–24 kbps). \textit{USAC-2} and \textit{USAC-3} use stereo signals at low (16–24 kbps) and high (32–96 kbps) bitrates, respectively.

The SEBASS-DB is a collection of subjective quality ratings for separated audio signal mixtures. It aggregates the results of five listening tests that assessed the basic audio quality of signals submitted to community-based source separation campaigns. All ratings were obtained using the MUSHRA test methodology. The database comprises four subsets — \textit{SiSEC08}, \textit{SASSEC}, \textit{PEASS}, and \textit{SAOC} — the last of which additionally includes signals rendered with the MPEG Spatial Audio Object Coding (SAOC) system.

\subsection{Results on Different Datasets}
In Tab.~\ref{TAB:eval_all}, we provide Pearson and Spearman correlation of the proposed system and other \ac{SOTA} system ratings against human ratings on the described datasets. 

The SCOREQ results are from no-reference (\textit{nr}) mode where the degraded signal is rated without any clean reference and from \textit{ref} mode where the results are the L2 distance of the embeddings of degraded and clean signals.
Please note that this evaluation procedure is different from the method presented in the original paper, where the authors use a non-matching reference.

Analyzing the results of Tab.~\ref{TAB:eval_all} shows that SCOREQ-\textit{nr} is the best performing system on all speech-only i.e. NISQA datasets but is failing to rate general audio content. When operating in \textit{ref} mode, however, SCOREQ produces meaningful ratings for speech as well as general audio, albeit with a surprising drop in performance on every speech dataset.
The PEAQ standard performs well on general audio content, but does not give strong correlation on speech.
DeePAQ performs very well on all datasets containing music, but shows limited performance on speech. This is likely due to the fact that this model is not trained on the NISQA dataset.

The proposed system achieves good results on speech as well as on general audio. However, its performance is generally limited on datasets that contain non-waveform-preserving items, such as ODAQ, USAC1, and USAC2.
This can be linked to the intrinsic waveform-dependent nature of the PEAQ precursors and can't be conquered by better models for representation learning or training strategies.
In both, SCOREQ and DeePAQ, the features of the underlying foundation model are averaged over time before calculating the embeddings and the final mapping to quality scores (trained either with or without reference). This makes these models independent of precise shape of waveforms and delay, which is also reflected in their good performance on these datasets.
\begin{table*}
    \centering
    \resizebox{\textwidth}{!}{%
        \begin{tabular}{c|ccccccccccccc}
        \toprule
           & NISQA-Live & NISQA-Val & NISQA-FOR & NISQA-501 & USAC1 & USAC2 & USAC3 & ODAQ & IGORC96 & SiSEC08 & SASSEC & PEASS & SAOC\\
           \midrule
          proposed & 0.71 / 0.73 & 0.84 / 0.81 & 0.71 / 0.72 & 0.75 / 0.71 & 0.73 / 0.72 & 0.71 / 0.69 & 0.84 / 0.87 & 0.77 / 0.82 & 0.89 / 0.78 & 0.9 / 0.84 & 0.87 / 0.8 & 0.79 / 0.63 & 0.88 / 0.73 \\
          PEAQ-2f & 0.53 / 0.53 & 0.74 / 0.79 & 0.66 / 0.66 & 0.75 / 0.77 & 0.86 / 0.87 & 0.79 / 0.69 & 0.91 / 0.94 & 0.87 / 0.82 & 0.94 / 0.86 & 0.96 / 0.87 & 0.94 / 0.89 & 0.87 / 0.55 & 0.95 / 0.86 \\
          SCOREQ-ref & 0.76 / 0.76 & 0.89 / 0.89 & 0.89 / 0.89 & 0.94 / 0.94 & 0.75 / 0.72 & 0.7 / 0.6 & 0.8 / 0.85 & 0.77 / 0.83 & 0.7 / 0.71 & 0.81 / 0.75 & 0.84 / 0.77 & 0.79 / 0.6 & 0.82 / 0.77 \\
          SCOREQ-nr & 0.79 / 0.79 & 0.92 / 0.92 & 0.93 / 0.93 & 0.94 / 0.95 & 0.14 / 0.14 & 0.14 / 0.14 & 0.09 / 0.09 & 0.02 / -0.03 & 0.06 / 0.01 & 0.54 / 0.55 & 0.63 /0.7 & 0.58 / 0.52 & 0.63 / 0.59 \\
          DeePAQ & 0.56 / 0.55 & 0.65 / 0.64 & 0.7 / 0.73 & 0.7 / 0.63 & 0.89 / 0.88 & 0.85 / 0.8 & 0.93 / 0.95 & 0.91 / 0.88 & 0.95 / 0.84 & 0.94 / 0.79 & 0.92 / 0.87 & 0.85 / 0.48 & 0.94 / 0.81 \\
          \bottomrule
        \end{tabular}
    } 
    \caption{Pearson and Spearman correlation on different test sets}
    \label{TAB:eval_all}
\end{table*}
\subsection{Explaining the Model's Predictions}
\begin{figure}[h!]
\centering
\begin{subfigure}{0.24\linewidth}
  \includegraphics[width=\linewidth]{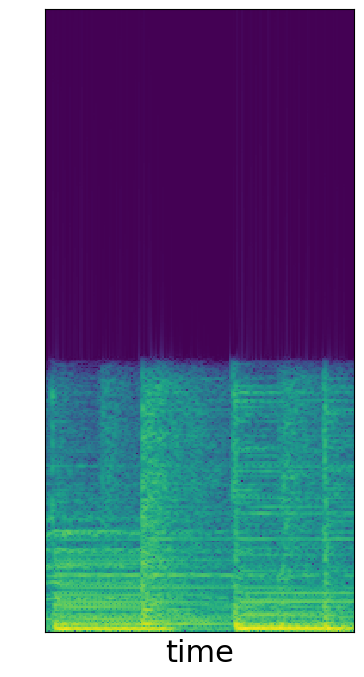}
  \caption{jazz}
\end{subfigure}
\hspace{-1mm}
\hfill
\begin{subfigure}{0.24\linewidth}
  \includegraphics[width=\linewidth]{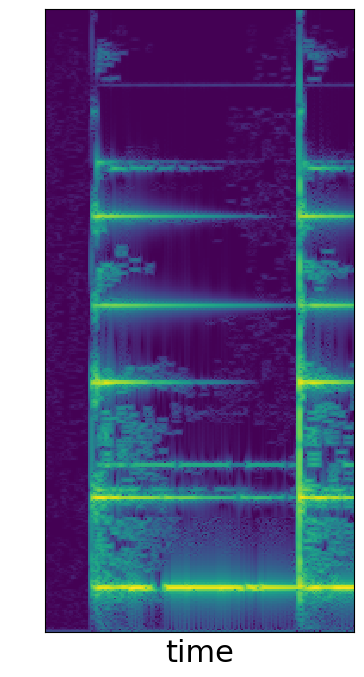}
  \caption{glockenspiel}
\end{subfigure}
\hspace{-1mm}
\hfill
\begin{subfigure}{0.24\linewidth}
  \includegraphics[width=\linewidth]{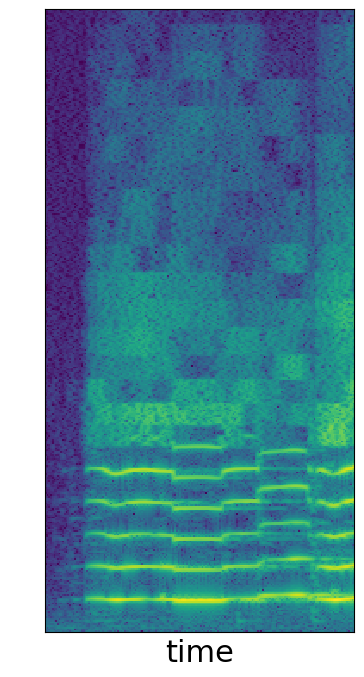}
  \caption{violin}
\end{subfigure}
\hspace{-1mm}
\hfill
\begin{subfigure}{0.24\linewidth}
  \includegraphics[width=\linewidth]{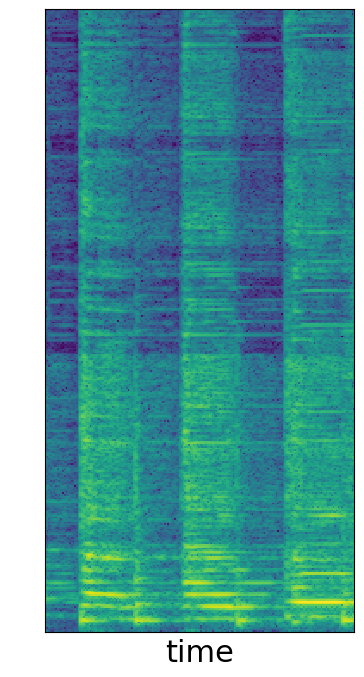}
  \caption{creatures}
\end{subfigure}
\vspace{-1mm}
\begin{subfigure}{0.24\linewidth}
  \includegraphics[width=\linewidth]{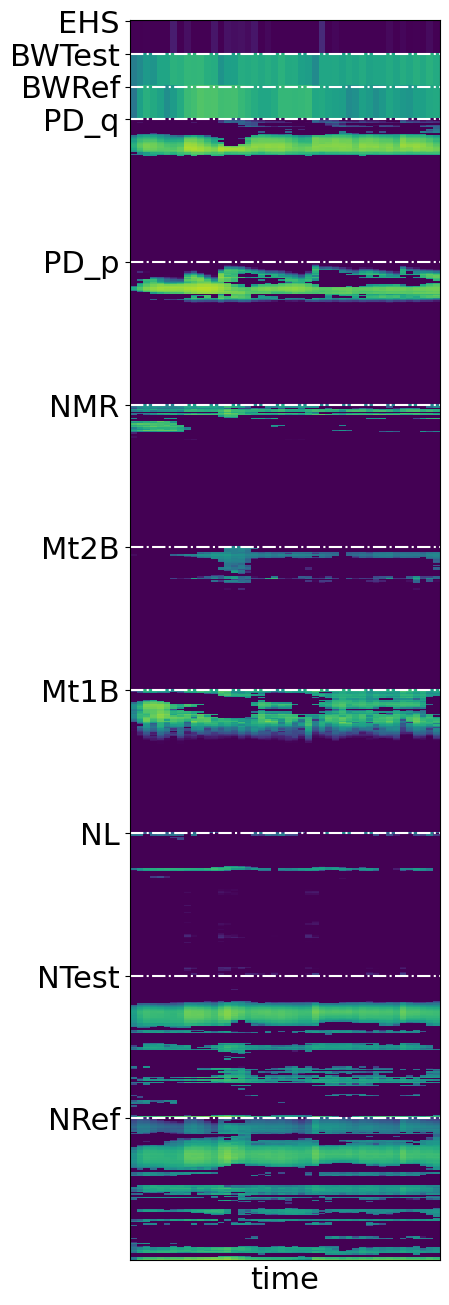}
  \caption{LP}
\end{subfigure}
\hspace{-1mm}
\hfill
\begin{subfigure}{0.24\linewidth}
  \includegraphics[width=\linewidth]{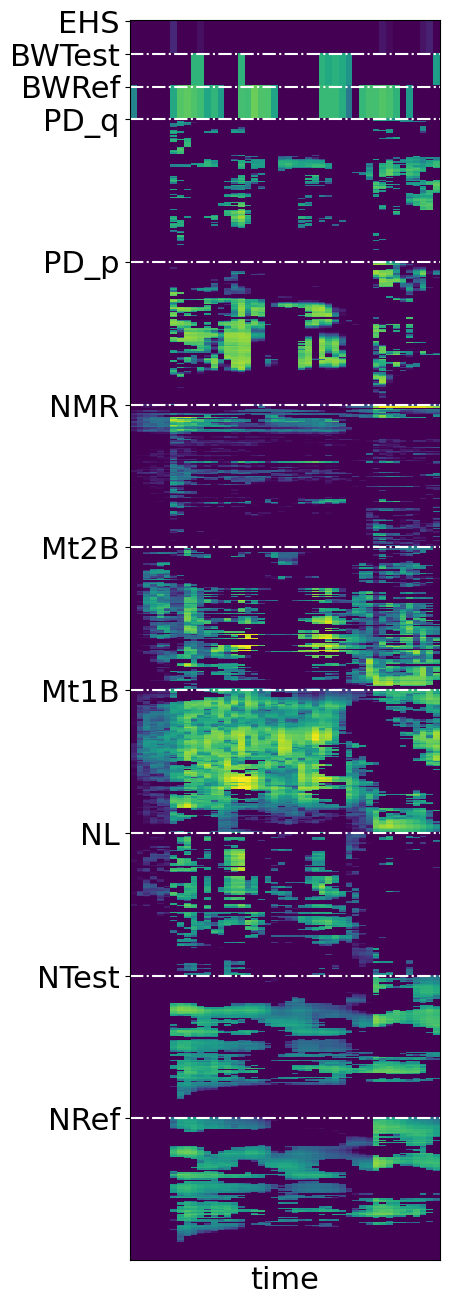}
  \caption{SH}
\end{subfigure}
\hspace{-1mm}
\hfill
\begin{subfigure}{0.24\linewidth}
  \includegraphics[width=\linewidth]{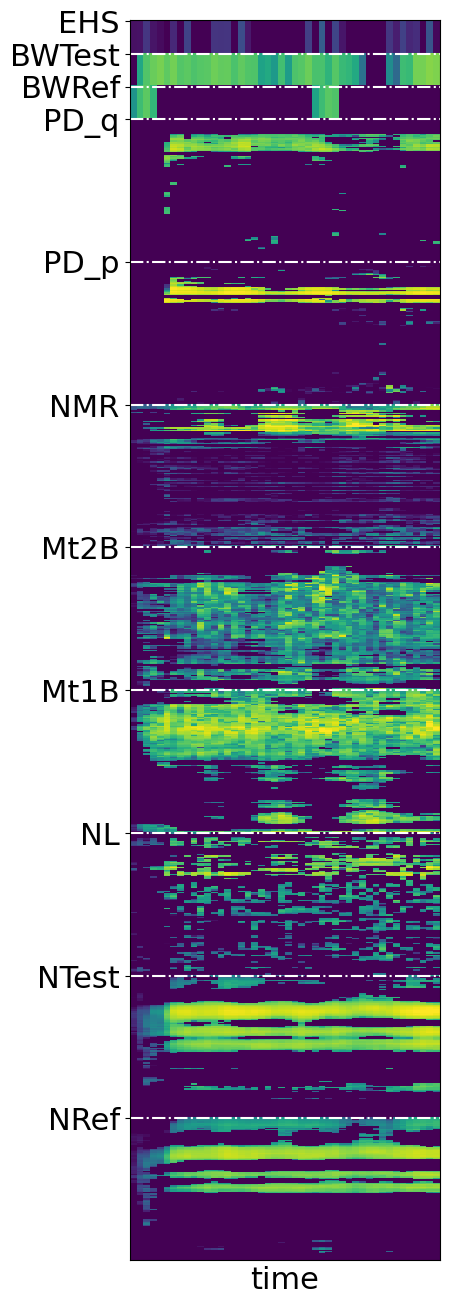}
  \caption{UN}
\end{subfigure}
\hspace{-1mm}
\hfill
\begin{subfigure}{0.24\linewidth}
  \includegraphics[width=\linewidth]{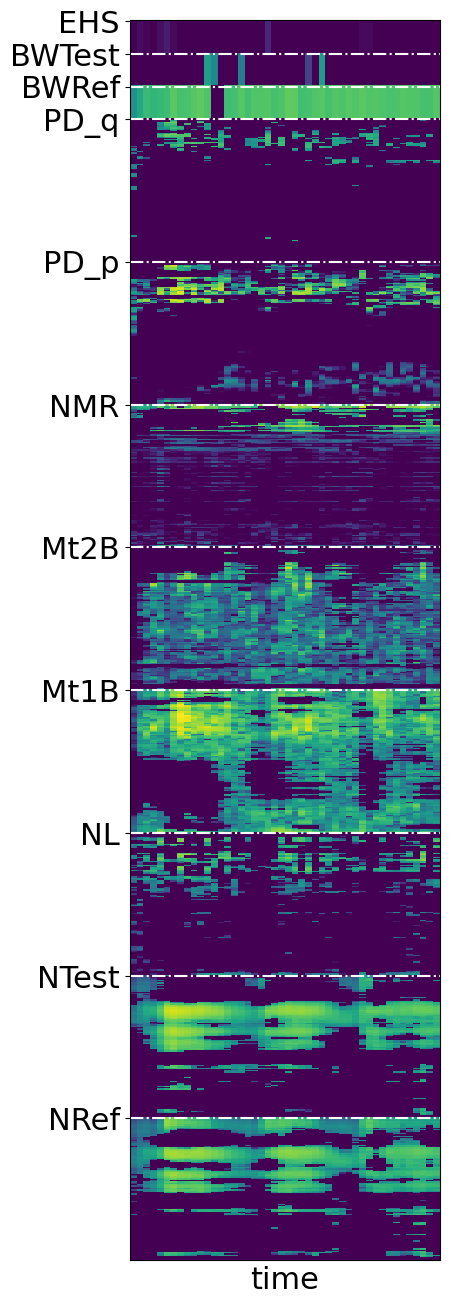}
  \caption{TM}
\end{subfigure}
\vspace{2mm}
\caption{Spectrograms (top row) and activation maps of precursors (bottom row) for 4 different degradation types from the ODAQ dataset.}
\label{fig:heatmap}
\end{figure}
\begin{figure}[h!]
  \centering
  \begin{minipage}{0.48\linewidth}
    \centering
    \includegraphics[width=\linewidth]{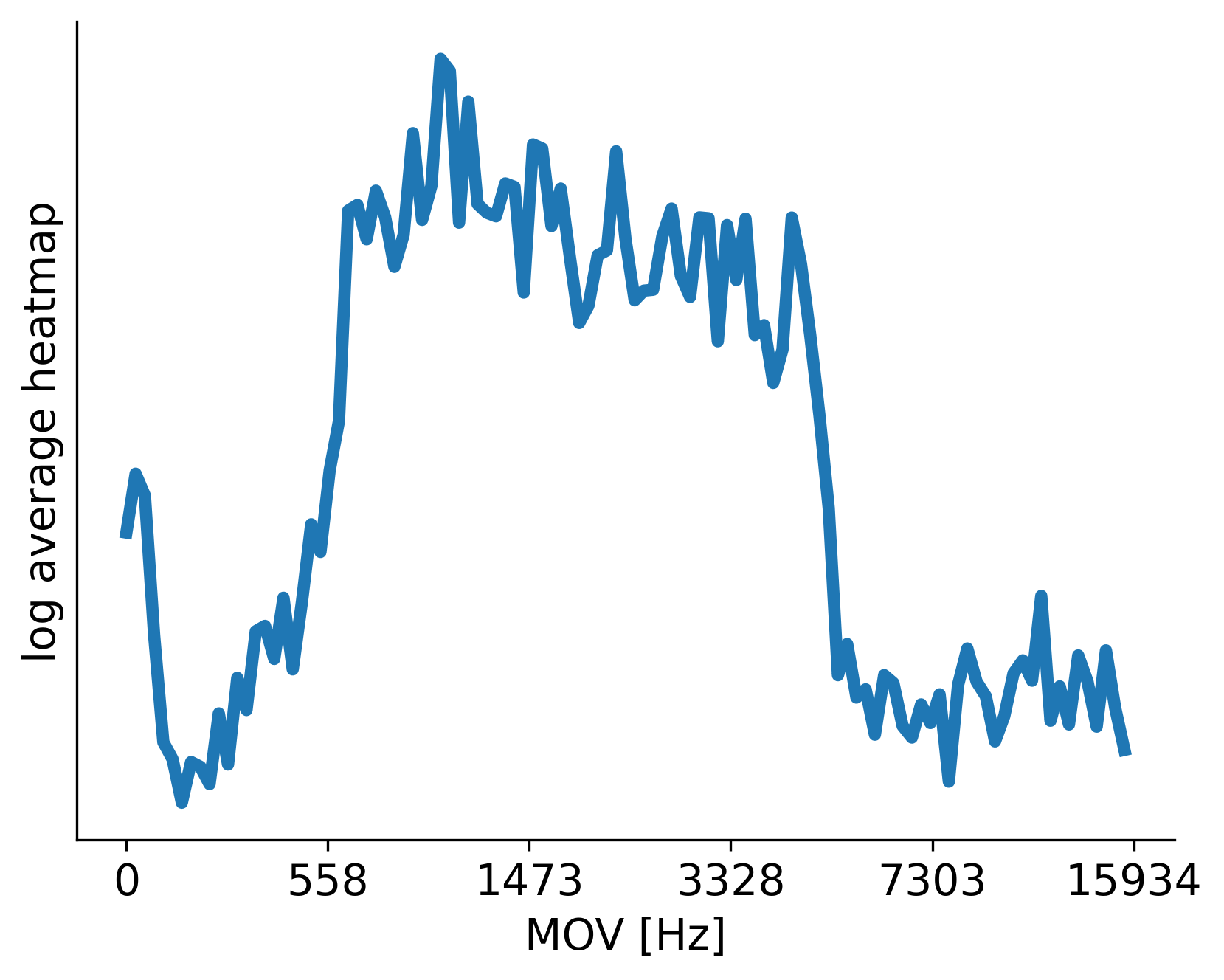}
    \vspace{-1mm}
    \caption{Heatmap averaged over time.}
    \label{fig:avg_time}
  \end{minipage}
  \hfill
  \begin{minipage}{0.48\linewidth}
    \centering
    \includegraphics[width=\linewidth]{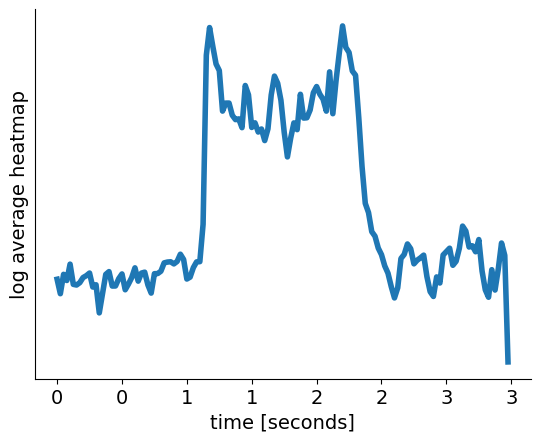}
    \vspace{-1mm}
    \caption{Heatmap averaged over frequency.}
    \label{fig:avg_frq}
  \end{minipage}
\end{figure}
In order to identify the source of the degradations, we apply the explainable AI method Grad-CAM.
It is a technique for visual \cite{GRAD_CAM} or auditory \cite{Expl_Acous_Scene_Class} explaining outputs of DNNs without requiring architectural modifications or re-training. 
It utilizes gradient information that travels backward to a specific layer to assess the significance of each neuron's contribution to a particular decision on the network outcome.
Originally designed for CNNs that do classification on images, we use it for our regression network by back-propagating loss $L$ to compute the gradient with respect to the desired layer
\begin{equation}
    L = \left\vert 5 - \mathrm{}e(\mathtt{pq}(\cleanSignal, \degradedSignal)) \right \vert.
\end{equation}
$L$ can be seen as a measure of how much the signal deviates from a clean reference (5 in our case).
This allows us to highlight important regions of the input features that contribute most to the models performance.
The selection of the feature map is not predetermined and subject to discussion but often the last convolutional layer of the DNN is chosen. However, since the input of our network is a set of features measuring auditory degradations, we chose the first convolutional layer as target feature map.

To assess whether Grad-CAM provides meaningful insight into the type of disturbance, we show the activation maps for four files from the ODAQ dataset in Fig. \ref{fig:heatmap}: lowpass (LP) applied to the file “jazz” (a); quantization with spectral holes (SH) applied to “glockenspiel” (b); unmasked high-frequency noise (UN) applied to “violin” (c); and tonality mismatch (TM) applied to “CreatureFromTheBlackjackTable”, abbreviated as “creatures” (d).
The dedicated heatmaps showing the activation of the precursors are shown in Fig.~\ref{fig:heatmap}.
Please note that all precursors except for \textit{EHS}, \textit{BWRef} and \textit{BWTest} have a frequency resolution on the y-axis in bark scale.

Analyzing the heatmaps (e) to (h), it becomes apparent that the precursors \textit{NRef} and \textit{NTest}, which measure the total loudness of the reference and the test signal per band, are almost always among the most active for all items. This suggests that our model extracts general information from them.

For the LP item, the most active precursors are, as expected, \textit{BWRef} and \textit{BWTest}, which measure the bandwidth of the reference and the test signal. 

For the SH item, Grad-CAM assigns the strongest responses to \textit{Mt1B} and \textit{Mt2B}. This is plausible because quantization-induced spectral holes are typically time-varying: spectral components disappear and reappear across frames, perturbing the temporal modulation of auditory-band excitation relative to the reference. The precursors \textit{NL}, which estimates the partial loudness of audible distortions, and \textit{NMR}, which relates error energy to the masking threshold, are also active. Their lower attribution is not necessarily unexpected, since omitted components may be partly masked, while their frame-wise variation can still produce strong modulation differences. For this item, the model therefore appears to rely more strongly on the temporal effects of the spectral holes than on their instantaneous error energy.

For the UN item, most precursors receive attribution. As expected, \textit{NL}, which estimates the loudness of audible disturbances, is prominent. The stronger attribution to \textit{Mt1B} and \textit{Mt2B} may arise because noise perturbs auditory-band envelopes and their temporal modulation.

The TM item produces a similar but weaker pattern, consistent with its higher rating (42 versus 26 MUSHRA points), provided that the maps share the same scale. Although \textit{EHS} is designed to capture harmonic structure in the error, it receives little attribution for this item.

Finally, we explicitly test temporal and spectral localization using two controlled 3-s items with deliberately localized distortions caused by AAC coding. In the first, degradation is applied only from 1 to 2\,s; in the second, it is restricted to 1.1--7.5\,kHz. Figures~\ref{fig:avg_time} and~\ref{fig:avg_frq} show the corresponding marginal attribution maps. In both cases, attribution is concentrated in the deliberately degraded region, confirming localization in time and frequency.










\section{Conclusion}
The main contributions of this work are first pretraining strategies that enable the usage of unlabeled data for training and second that we use Grad-Cam to make the model interpretable. The latter is particularly interesting when using features that measure certain auditory events like the PEAQ features. Hence, our model is able to give feedback to the user about the kind, locality and amount of distortion present.
We also show strong correlation with human ratings on a large variety of data, including speech and general audio
\clearpage

\bibliographystyle{IEEEbib}
\bibliography{refs_cleaned}

\end{document}